\documentclass[prd,aps,preprintnumbers,twocolumn,floatfix]{revtex4}
\usepackage{epsfig}
\usepackage{amsmath}

\def\f{{\rm f}}

\def\bea {\begin{eqnarray}}
\def\eea {\end{eqnarray}}

\begin{document}

\preprint{YITP-SB-08-45}

\title{Soft-Gluon Cancellation, Phases and Factorization
with Initial-State Partons}

\author{S.\ Mert Aybat$^{1}$ and George Sterman$^2$}

\affiliation{$^{1}$Institute for Theoretical Physics, ETH Zurich,
 8093 Zurich, Switzerland}
\affiliation{${}^2$C.N.\ Yang Institute for Theoretical Physics, 
Stony Brook University, Stony Brook, New York 11794--3840, USA}
\date{\today}

\begin{abstract}
We outline arguments for the cancellation of soft singularities in transition
probabilities and parton distributions with incoming partons and Wilson lines,
and with  observed
jets  or heavy colored particles in the final states.
We  show the cancellation of Glauber gluons
and divergent phases, 
relating finite remainders to
corrections that arise from
restrictions on final states in factorized cross sections.
\end{abstract}

\maketitle
In this paper we 
investigate
the cancellation of 
soft-gluon singularities
for inclusive 
transition probabilities with a parton and a Wilson line in the initial state.   
Our arguments apply to a large class of parton distributions, and 
outline an extension of the proof of  factorization in Ref.\ \cite{Collins:1988ig}
for Drell-Yan and related processes 
\cite{Collins:1981ta,Frenkel:1983di,Bodwin:1984hc} to
 cross sections with colored particles in the initial state and with final-state jets.
This requires us 
to explore the role of ``Glauber" or ``Coulomb" regions of momentum
space~\cite{Collins:1981ta,Bodwin:1981fv,Collins:1983ju, Collins:2004nx} in these processes.   These
 regions, where soft gluons carry unphysical spacelike momenta, 
 are related to the generation of phases \cite{Forshaw:2006fk,Collins:2007nk}. 

{\bf Transition Probabilities.}
We consider 
weighted transition probabilities with
an initial-state Wilson line in color representation $f'$ and a parton of flavor $f$ and momentum $p$,
\vspace{-3mm}
\begin{eqnarray}
{\cal S}_{\f}[S]&=&\sum_N
S(N) 
{\rm Tr}_c\ \Bigg[\;
\big \langle p| \bar T \left[  \left ( \phi_f(0) \Phi^{(f')}_\beta{}(0,-\infty) \right)^\dag \right ]  
|N \big \rangle\nonumber\\
&\ &\hspace{-3mm}\times
\big \langle N|\, T\left ( \, \phi^\dagger_f(0)\,\Phi^{(f')}_\beta(0,-\infty)\,  \right ) |p \big \rangle\
\Bigg]\, ,
\label{eiksig}
\end{eqnarray}
where $T$ and $\bar{T}$ denote time- and anti-time-orderings and $\f\equiv \{f,f'\}$. 
The field of the incoming parton is $\phi_f$,  $\beta^\mu=\delta_{\mu +}$ is a light-like four-velocity 
in the plus direction,
and  
$
\Phi_{\beta}^{(f')}(b,a)
= P\ \exp \left[ -ig\int_{a}^b d\lambda\,   A_\beta(\lambda\beta)\right]$ 
is the Wilson line, with $f'$ the group representation for the gauge field
$A_\beta\equiv \beta\cdot A$.   We take $p^\mu=p^-\delta_{\mu -}$ opposite to $\beta$. 
The function $S(N)$ assigns weights to final states $|N \rangle$.   
For our arguments we will 
assume
that $S(N)$ 
satisfies the usual requirements for infrared safety 
\cite{Sterman:1978bi}, and depends implicitly on a hard scale $Q$.     
It may be chosen, for example,
to select final states with heavy particles and/or  jets with  specified energies and directions.
The weight function $S(N)$ may be also chosen to fix the
total
 transverse momentum
of the final state.   
The connection between jet and single-particle inclusive cross sections with hadronic initial states was recently reviewed in Ref.\ \cite{Nayak:2005rt}.
Defining $\phi_f$ with an appropriate spin projection, the
class of probabilities ${\cal S}_{\rm f}$ includes a variety of 
gauge-invariant \cite{Collins:1981uw}
spin-dependent parton distributions
with incoming Wilson lines \cite{Collins:2004nx,Collins:2002kn}.
The all-orders cancellation of soft singularities in such distributions is to our knowledge a new result.
The direct relation to factorization in hadron-hadron scattering is through the ``weak factorization"  of
Refs.~\cite{Collins:1981ta,Bodwin:1984hc}.   

{\bf Cancellation for the Target Field.}
Expanding the Wilson lines of (\ref{eiksig})  and writing 
the expansion in terms of momentum space integrals of Fourier-transformed fields gives
\vspace{-1mm}
\begin{eqnarray}
{\cal S}_{\f}[S] &=& 
\sum_{n=0}^\infty g^n  
\prod_{i=1}^n \int d^4q_i \sum_{j=1}^n E^{(f')}{}^n_j(\beta\cdot q_i)
 \nonumber\\
 &\ & \hspace{-15mm} \times \sum_N S(N)
 {\rm Tr}_c \left[
\big \langle p|\,   A_\beta(q_{j+1}) \times \cdots \times A_\beta(q_n)  \phi_f^\dag(0)\, 
 |N \big \rangle\ \right.
 \nonumber\\
 &\ & \hspace{10mm} \left. \times\
 \big \langle N|\, \phi_f(0)\,
  A_\beta(q_1) \times \cdots \times A_\beta(q_{j})\, |p \big \rangle\; \right]
  \nonumber\\
  &\equiv& \sum_{n,j}\, E^{(f')}{}^n_j\,\otimes_q\, \sum_N R_f^{(N,n,j)}[S]\, ,
\label{eiksigA(x)b}
\end{eqnarray}
 where  $q_i$ is the momentum flowing out of
field $A(\beta\lambda_i)$, and all dependence on final states is in $R_f^{(N,n,j)}$. 
In covariant gauges,
soft-gluon singularities arise only from regions where sets of the  $q_i$ vanish \cite{Libby:1978qf}.
The position-space integrals in the ordered exponentials of Eq.\ (\ref{eiksig}) 
give  eikonal 
$\beta\cdot q$-dependent factors, independent of final state $N$,
\begin{eqnarray}
E^{(f')}{}^n_j(q_1^-,\dots,q_m^-)&=&\Big(\frac{1}{q_{j+1}^-+i\epsilon}\,\frac{1}{q_{j+1}^-+q_{j+2}^-+i\epsilon}
\cdots \nonumber\\
&\ & \hspace{-20mm}\times
\frac{1}{q_{j+1}^-+\dots+q_n^-+i\epsilon} \Big) (-1)^{n-j}\\
&\ &\hspace{-34mm} \times   \Big( \frac{1}{q_1^-+\cdots+q_{j}^-+i\epsilon} 
\cdots 
\frac{1}{q_{j-1}^-+q_{j}^-+i\epsilon}\frac{1}{q_{j}^-+i\epsilon}
\Big),\nonumber
\label{eikdenom_a}
\end{eqnarray} 
where 
$q^-\equiv \beta\cdot q$. The $E^{(f')}{}^n_j$ satisfy \cite{Libby:1978nr,Bodwin:1984hc}
 \begin{equation}
 \label{Eq:identty_a}
\sum_{j=0}^n\,E^{(f')}{}^n_j(q_1^-,\dots,q_n^-)=0\, ,
\end{equation}
a result 
we will   
refer to as the ``eikonal identity" below.
First, however, we must discuss the functions $R$,
recalling the analysis of Ref.\ \cite{Collins:1988ig}.

In Ref.~\cite{Collins:1988ig}  the remainder functions $R_f^{(N,n,j)}$ 
were treated in light cone-ordered perturbation theory (LCOPT), by integrating over the
plus momenta of their internal loops at fixed minus and transverse momenta, 
including an integral over the total plus momentum flowing
from the Wilson line into the final state, $p_N^+$.   
In this formalism, the cancellation of final state interactions 
\cite{Collins:1983ju} is manifest after the integral over
$p_N^+$, and one finds that for fixed $n$, the result is actually independent of $j$, 
that is, of how the set of gluons is split between the left and right eikonals,
\vspace{-1mm}
\begin{eqnarray}
\sum_N R_f^{(N,n,j)}&=& \mathcal{P}^{n}[S,q_i^-] \, .
\label{sumRf}
\eea
Corrections vanish for a fully inclusive sum, $S(N)\equiv 1$, and are otherwise infrared finite.
The function ${\cal P}^n$ is given by 
purely initial-state LCOPT factors,
\bea
\mathcal{P}^{n}[S,q_i^-] &=& \sum_N S(N)\,
\sum_{T\in {\cal T}_I} \int_{k_l}\,I_T^{(N)\,'*}(k_l)\,  I_T^{(N)}(k_l)
\nonumber\\
&\ & \hspace{5mm} \times \ \prod_{i=1}^n\, \theta\left( \delta_i^{(T)} q_i^-\right)\, ,
\label{sumNofR}
\end{eqnarray}
where  the sum is over
those orderings $T$ in which every vertex precedes the annihilation of the incoming parton,
and where
the integrals are over internal minus and transverse loop momenta, $k_l$.
For each light-cone ordering $T$, $\delta_i^{(T)}=\pm 1$ fixes the flow of $q_i^-$
from past to future,
independent of $j$.
The functions $I_T$ are products of LCOPT denominators, 
given by the plus momentum deficit of each 
initial
state, $\xi$, relative to the incoming 
state with $p^+=0$,
\bea
I_T^{(N)}(k_l)=\prod_{\xi <H}\frac{1}{-s_\xi + i\epsilon}\, ,\;
s_\xi \equiv \sum_{lines\, j\  j\in\xi} \frac{k_{j,\perp}^2}{2k_j^-}\, ,
\eea
where $<$ refers to earlier in light-cone ordering, and where
we suppress numerator factors.   We denote as $H$ the
vertex at which the parton $f$ is annihilated, to divide initial from final states.
 Using Eqs.\ (\ref{Eq:identty_a}) and (\ref{sumRf}) in 
 Eq.\ (\ref{eiksigA(x)b}) , we conclude that, when all the $q_i^-$ are nonzero,
the sum over all connections of gluons to the initial-state Wilson line vanishes,
and soft-gluon singularities cancel.

{\bf The Eikonal Field and Induction.}
By itself, the forgoing would appear 
 to eliminate all gluon attachments to the eikonal lines
and rule out all soft-gluon singularities in Eq.\ (\ref{eiksig}).
This argument has an important limitation, however,
related to the output of LCOPT, Eq.\ (\ref{sumNofR}).   This is 
an ambiguity in
the
integrations at points $q_i^-=0$ where, in general, the eikonal factors are singular.
On the one hand, the eikonal singularity at $q^-=0$ for a virtual gluon exchanged between
the incoming Wilson line and the incoming parton $f$ is clearly related to
a divergent imaginary ``phase". 
On the other hand, in
order to use  the eikonal identity (\ref{Eq:identty_a}), we must sum over final
states with gluons whose plus momenta are given by $q_i^+=q_{i,T}^2/2q_i^-$.
In any realistic case, there is a maximum value, $Q$, on the $q_i^+$
and hence a lower limit on $q_i^-$ when $q_i$ appears in the final state.
We shall refer to
partons whose minus components are large enough to satisfy this
bound the {\it target field}.
The cancellation of imaginary parts is 
evidently not included in the
eikonal identity of Eq.~(\ref{Eq:identty_a}), which applies to the target field only. 

In summary, to establish cancellation we must extend 
the analysis of Ref.\ \cite{Collins:1988ig} to regions of arbitrarily low $q^-$.
We will refer to the set of low-$q^-$ lines connected to the incoming eikonal line as the 
{\it eikonal field subdiagram}, ${\cal E}^{(f')}$.
 In the eikonal field subdiagram, the plus momentum integrals 
 that lead to LCOPT expansions do not generally
 converge \cite{Collins:1988ig}, which leads precisely to the ambiguity
 in the minus integrals noted above.   We shall treat this problem below.
 
In general, lines of the eikonal field subdiagram ${\cal E}^{(f')}$ may appear in the
final state carrying plus momenta of order $Q$.   Such lines are
``spectator lines" of a jet in the $\beta$ direction.   
Adapting
the 
reasoning
of Ref.\ \cite{Collins:1988ig}, one can
show that $\beta$ jet spectators factorize 
from soft gluons after a sum over final states, 
and hence do not participate in soft divergences.   
We will assume this result here, and review details of the 
extended 
argument elsewhere \cite{AS}.
In this paper, we will show cancellation for
transition probability ${\cal S}_{\rm f}^{({\cal E})}$,
in which the eikonal field subdiagram is entirely virtual.
Our argument for ${\cal S}_{\rm f}^{({\cal E})}$ 
is inductive, with the cancellation for the pure target field, described 
 above, as the inductive basis.  Induction
is carried out in the loop order of
the (virtual) eikonal field subdiagram, ${\cal E}^{(f')}$.
In this transition probability, we must sum over only those states for which
${\cal E}^{(f')} = {\cal E}^{(f')}_{M^*} \oplus {\cal E}^{(f')}_{M}$, a sum of independent
subdiagrams of the amplitude and its complex conjugate.
The on-shell plus momenta of all final state particles is then
bounded by the hard scale $Q$.   We 
denote these ``target field" final states by $N_T$.

An IR safe weight function $S(N)$ can have only smooth dependence
on target field states, and for most of the remaining analysis we set $S(N)=1$,
and show how infrared divergences associated with 
these final states cancel.   We will then return to the choice of $S(N)$.
In the discussion below, we will also restrict
ourselves to regions in momentum space where all transverse
momenta are of a comparable size, which we denote by $m$.   We
will see below that extensions to multiple scales are
 intimately connected with the choice of the weight $S(N)$.  For now,
 $m^2/Q$ represents a lower limit for the minus momenta in the target field.

{\bf The Hybrid Form.}
To discuss the role of the eikonal field,
we express contributions to the transition amplitude for fixed 
eikonal connections $n$ and $j$ in
Eq.\ (\ref{eiksig}), summed over target field
final states $N_T$, as
\vspace{-1mm}
\begin{eqnarray}
\sum_{N_T}\, E^{(f')}{}^n_j\otimes R_f^{(N_T,n,j)}[1,q_i^-] &=&
\nonumber\\
&\ & \hspace{-30mm}
\sum_{N_T}\, \int d\tau_{N_T} \,  M^*_{n-j}(p_t)\, M_j(p_t)\, ,
\label{ERtoMM}
\eea
where $p_t$ denotes the momenta of  lines $t$ in the final state,
and $d\tau_{N_T}$ their phase space measure.   Once again, final state
interactions cancel in the sum over 
$N_T$.

Consider now a graphical contribution to the amplitude $M_j(p_t)$
in Eq.\ (\ref{ERtoMM}).   In general, it
is a combination of an eikonal subdiagram ${\cal E}_M^{(f')}$
and a ``reduced remainder", $r_{{\cal E}^{(f')}} \equiv M_j/{\cal E}_M^{(f')}$, none of 
whose lines have minus momenta of order $m^2/Q$ or less.
In general, the plus momenta of lines $q_i$ of ${\cal E}_M^{(f')}$ that attach to $r_{{\cal E}^{(f')}}$
are ``pinched" between poles at 
$q_i^+ \sim \pm (m^2 - i\epsilon)/Q$ from active and spectator lines of
 $r_{{\cal E}^{(f')}}$ in covariant perturbation theory \cite{Collins:1981ta,Bodwin:1981fv}.
Such lines are often referred to as ``Glauber exchanges".
We now evaluate $r_{{\cal E}^{(f')}}$ in LCOPT, treating the eikonal field 
as an external source of gluons at fixed momenta, $q_i^+$.

LCOPT provides a sum of $x^-$-ordered diagrams
for $r_{{\cal E}^{(f')}}$.  
Suppressing numerator momenta, the amplitudes are, as above,
products of inverse plus momentum deficits
between
the total plus momentum that has flowed into the diagram before each state, $i$, and the
on-shell plus momentum, $s_i$ of all the particles in the state.
We refer to the resulting expression as a {\it hybrid} form for $M_j$,
\vspace{-1mm}
\bea
M_j(p_t) &=& \int_{q}{\cal E}^{(f')}_M(\{q_a,q_b\})\, \, 
\prod_{i<H} \frac{1}{ -\ \sum_{a \in i} q_a^+ - s_i+i\epsilon}
\nonumber\\
&\ & \times \prod_{j>H} \frac{1}{  \sum_{b \in j} q_b^+ + k^+_{N_T} - s_j + i\epsilon}\, .
\label{Mj}
\eea
The first (second) product of LCOPT denominators represents initial (final) states
relative to vertex $H$.
The function ${\cal E}^{(f')}_M(\{q_a,q_b\})$ denotes the eikonal field subdiagram, whose
low-$q^-$ external lines attach to $r_{{\cal E}^{(f')}}$ either in initial states ($q_a$)
or final states ($q_b$), with an integral over the allowed region for all loop momenta.
The $s_i$  
are sums of terms of order $m^2/Q$ for particles in
the $p$ jet, and order $m$ for soft particles.   
On-shell plus
momenta can be order $Q$ for
particles in outgoing jets, but  $k_{N_T}^+- s_j$ is
order $m$ or less.

{\bf Glauber Exchange and the Causal Identity.}
By construction, all lines of
${\cal E}^{(f')}$ 
have minus momenta
of order $m^2/Q$ and transverse momenta of order $m$; hence singularities of the $q_{a,b}^+$
from ${\cal E}^{(f')}$ in Eq.\ (\ref{Mj}) are  at order $Q$.   
In  addition, in
Eq.\ (\ref{Mj}), the $q_{a,b}^+$ encounter poles from $r_{{\cal E}^{(f')}}$ only in the  UHP
when  they enter 
$r_{{\cal E}^{(f')}}$
in an initial state, and only in the LHP
for a final state.    
As a result, we may deform the $q_{a,b}^+$ contours
to order $Q$ from the origin to complex, collinear-$\beta$ values.
This deformation is into the LHP for the $q_a$, flowing out of initial states,
and into the UHP for the $q_b$ and final states.   Along the
deformed contours, 
$k^+_{N_T}-s_j$ can be neglected in each term individually.  
The Glauber ``pinch" of 
covariant perturbation theory is then replaced
in the hybrid form
 by a mismatch between
the directions  of the deformations for initial and final states.
In addition, we see that the eikonal field gluons can appear
only with one denominator per integration if the result is
to be leading power in $Q$.    The only orderings that survive
are those in which, say, 
$h$ eikonal field gluons appear in all possible
combinations in $l$ initial and $h-l$ final states.   
Each such integral has logarithmic power counting overall.

\begin{figure}[h]
\centerline{\epsfxsize=2cm \epsffile{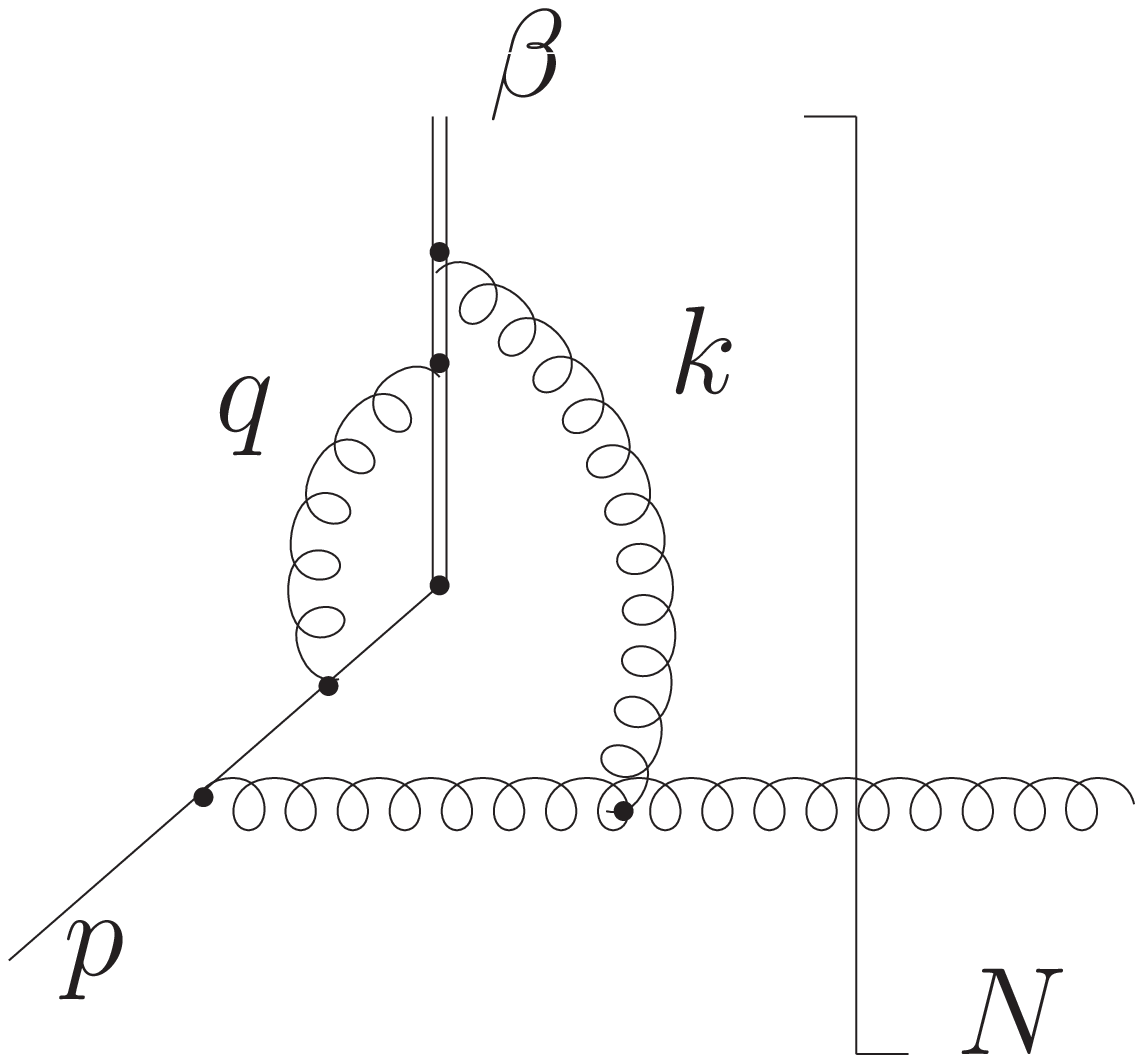}
\hspace{2mm} \epsfxsize=2cm \epsffile{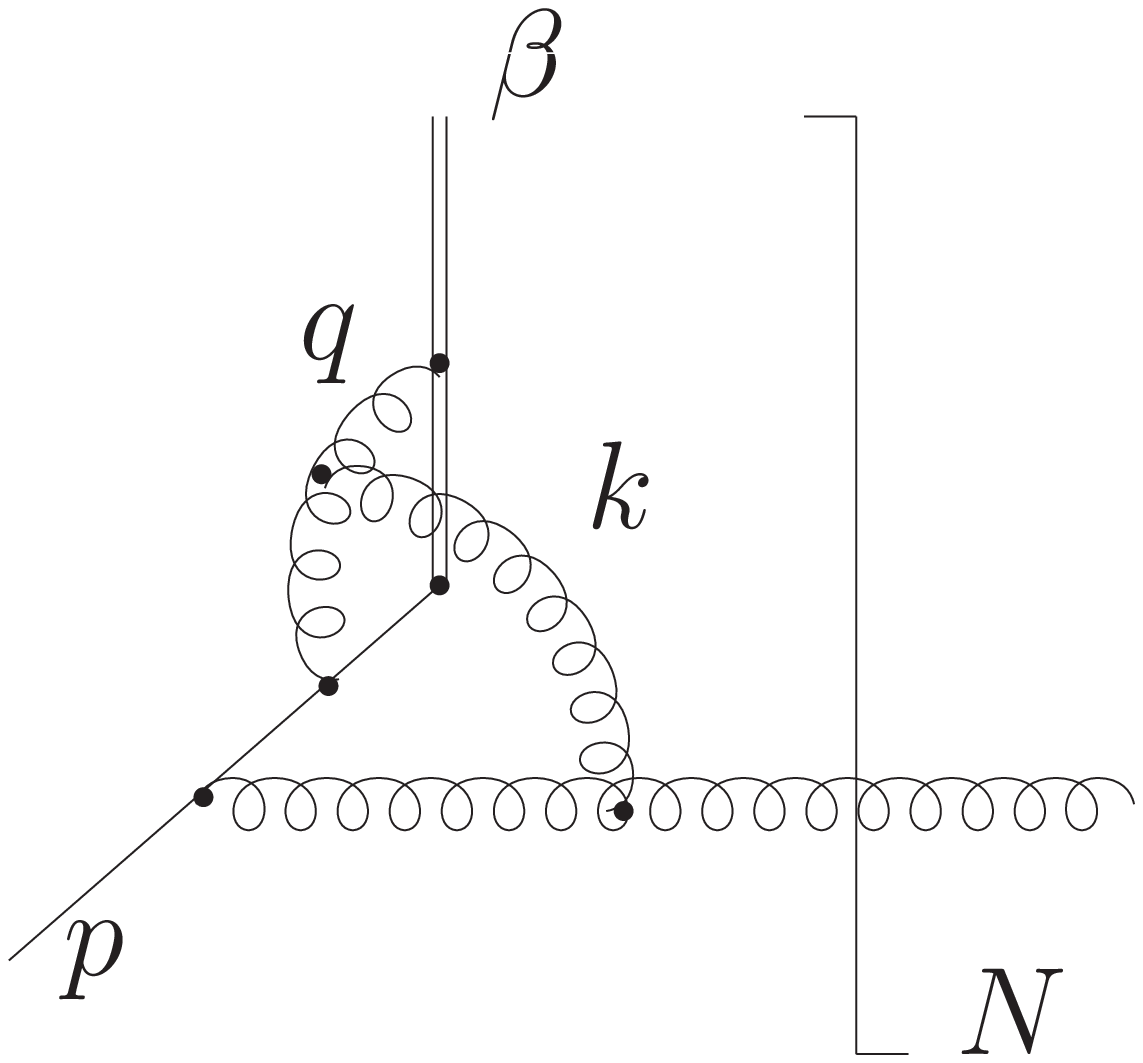}
\hspace{2mm}\epsfxsize=2cm \epsffile{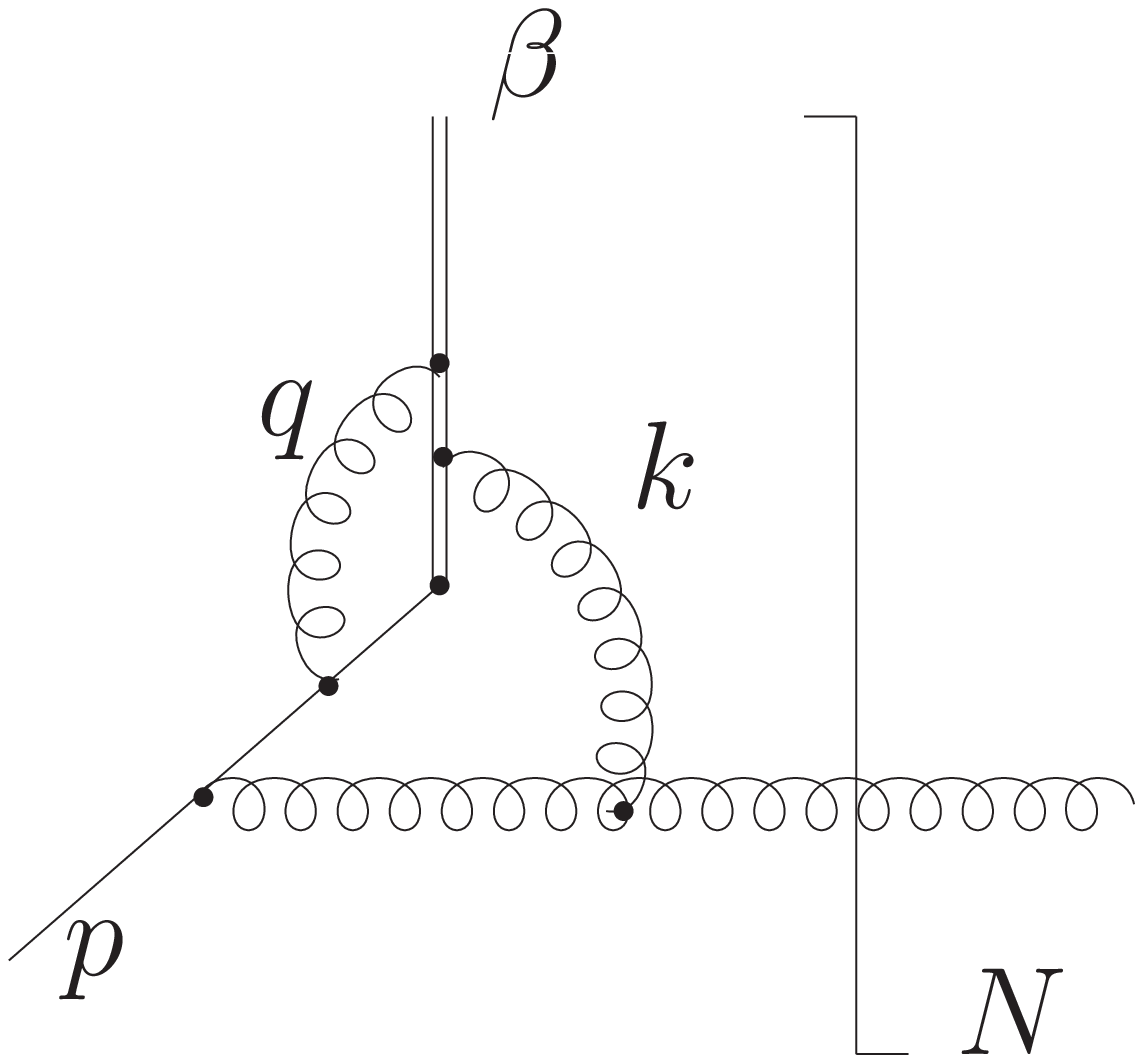}} \ \\
\hbox{\hskip 20mm (a) \hskip 20 mm (b) \hskip 18 mm (c)} 
\caption{Examples for active and spectator connected diagrams in LCOPT.
\label{figs}}
\vspace{-3mm}
\end{figure}
We now distinguish between ``active-connected" eikonal field
gluons that attach to ``active" lines in $r_{\cal E}^{(f')}$, whose minus 
momenta flow into vertex $H$, and ``spectator-connected" lines,
that attach to lines in $r_{{\cal E}^{(f')}}$
whose minus momenta flow into the final state.
Examples are shown in Fig.~\ref{figs}.
When active-connected eikonal field lines acquire large real
plus momenta, they are properly considered part of the
$\beta$ jet, and appropriate subtractions will
 ensure 
convergence of the corresponding $q_a^+$ integrals \cite{Bodwin:1984hc}.   
Spectator lines with large plus momentum
in the hybrid form are not naturally part of the $\beta$ jet;
indeed they are a reexpression of Glauber exchange.
No subtraction is necessary, however, because
 the sums over spectator attachments to initial and final states 
 cancel by the same identity as in Eq.\ (\ref{Eq:identty_a}),
\bea
 \sum_{a=0}^n
\prod_{k=1}^a \frac{1}{ - \sum_{i=1}^k q_i^+ }
\ \prod_{l=a+1}^n \frac{1}{  \sum_{j =l}^{n} q_{j}^+  } = 0\, ,
\label{causal}
\eea
which can be interpreted as the quantum-mechanical 
suppression of acausal correlations \cite{AS}.
Note that the ``causal identity" (\ref{causal}) holds for fixed color orderings in each diagram,
and whether or not observed jets are included in the final state.
The contributions of large-plus momentum gluons attached to spectators, 
although present in individual LCOPT diagrams, thus vanish in the sum.

In summary, once it is collinear-subtracted, $M_j$ will have no
contributions from large plus momenta.   At the same time, 
we cannot simply ignore the spectator-connected eikonal field, because the 
use of the causal identity assumes that we deform the $q_b^+$
and $q_a^+$ contours into {\it the same half-plane}.   This requires
us to cross poles from the LCOPT denominators in (\ref{Mj}), setting 
some $q_b^+$, for example, to values characteristic of the intermediate
states of the target field, of order $\le m$.   Similar
considerations apply to active-connected eikonal gluons, because
the subtractions for the $\beta$ jet should be defined with
respect to an outgoing rather than incoming Wilson line \cite{Bodwin:1984hc}. 
The poles, then, are the remaining effects of the divergent plus
integrals of the eikonal field, and are the origin of divergent
phases not present in the 
purely
target field integrals.

{\bf Wilson Lines and Spectator Interactions.}
To treat the pole contributions, we sum over the amplitudes $M_j$ for fixed final states, $N_T$.
As we have seen, the eikonal field momenta $q_{a,b}$ in
(\ref{Mj}) appear only in a minimum number
of initial or final states.   As a result, the number of spectators is
conserved among these states.    The initial- (final-) state
interactions of eikonal field  gluons with spectators 
then factorize from the spectator lines onto incoming (outgoing) Wilson
lines  of the corresponding color representations (labelled $t$ below),
 \cite{Collins:1988ig,Bodwin:1984hc,SCfact}.
Nonvanishing eikonal-field-spectator interactions are thus generated by the operators \cite{AS}
\bea
W_-^{(t)}({\bf x}_t) = \Phi^{(t)}(\infty,{\bf x}_t)\Phi^{(t)}({\bf x}_t,-\infty)\, .
\label{Wdef}
\eea
These Wilson lines are local with respect to transverse momenta,
which flow into the initial states of $M_j$, Eq.\ (\ref{Mj}).    We may then reorganize
the sum  ${\cal M}\equiv\sum_j M_j$ as
\vspace{-1mm}
\bea
&& {\cal M}(p_a,\{p_t\}) = 
 \Big \langle 0 \Big | \Phi^{(f')}(0,-\infty)\, C_{f'f}(p_a)\, \Phi^{(f)}(0,-\infty)
\nonumber\\
&& \hspace{-2mm} \times\ \prod_t \int d^2{\bf x}_t\,  e^{i{\bf p}_{t,\perp}\cdot {\bf x}_t}\, 
W_-^{(t)}({\bf x}_t)
\Big | 0 \Big \rangle \Psi_f(\{{p^-}_t,{\bf x}_t\} ))\, ,
\label{Mjlines}
\eea
with $C_{f'f}(p_a)$ 
a short-distance function
that links the incoming Wilson lines
in representations $f'$ and $f$.    
 The vertex $C_{f'f}(p_a)$ includes all information on the production of jets
 or heavy colored particles, whose interactions cancel in
 the sum over final states.   
 In general, $C_{f'f}(p_a)$
is a sum over terms in all color representations that arise from the direct product
of the two partonic representations of $f'$ and $f$.   
It depends on $\beta$ and
on the total ``active parton" momentum $p_a$, collinear to $p$,
that takes part in the hard scattering.
Because soft interactions between final-state jets and other particles
 cancel \cite{Libby:1978nr}, the color representation in $C_{f'f}(p_a)$ and 
in the corresponding factor in the complex conjugate amplitude is the same
for an inclusive cross section.

In Eq.\ (\ref{Mjlines}), $\Psi_f$ is a light-cone wave function  \cite{Brodsky:1997de} for
the incoming parton $p$, which incorporates all initial-state
dependence from states earlier than the earliest connection
of the eikonal field to the target field.    The wave function depends on 
$p$ and a
set of collinear momenta $p_t^-$ of the observed spectators, 
each at a fixed transverse position, ${\bf x}_t$.   Of course, $p=p_a+\sum_tp_t$.
In effect, we find that at the level of amplitudes,
Glauber gluons act to dress light-cone wave functions
with nonabelian phases.

We now treat  ${\cal M}$ in hybrid form.
The  hard-scattering vertex
 and the ``midpoints" of the spectator eikonals $W_-^{(t)}$ at $x={\bf x}_t$ 
 are local in $x^-$, 
  so that  plus momentum loops flow from the eikonals to the hard scattering
 in LCOPT at this vertex, just as in Eq.\ (\ref{Mj}).
 Now at fixed plus momenta $\le Q$, the only 
enhancements 
for minus momenta $\le m^2/Q$
are from the 
eikonal denominators.   Thus,  loops of ${\cal E}^{(f')}$ that do not flow directly
through the $\beta$ eikonal must carry plus momenta of order $Q$
to give leading contributions, as for momentum $k$ in the example of Fig.~\ref{figs}b. 
This means
that all loops of any eikonal field subdiagram that connects the
$\beta$ eikonal to more than one spectator or to 
the active line and one or more spectators must carry
large plus momentum into the $x^-=0$ vertex.   
Such contributions cancel
by the causal identity.

 {\bf Subtraction and Cancellation.}  
Purely collinear divergences will remain after soft gluon cancellations, and
are factorized from the transition amplitude into an ``eikonal" distribution function
\cite{Laenen:2000ij}
for partons collinear to $\beta$.
  We shall assume that this factorization can be implemented 
by collinear subtractions \cite{Collins:1988ig,Bodwin:1984hc}.

Following the inductive approach described above, 
we consider integrals
over minus momenta $q_i^-$ of ``active" loops, connecting the active eikonal line and
the $\beta$ eikonal in (\ref{Mjlines}) at fixed values of their plus momenta.
Any double-counted regions 
have smaller eikonal field subdiagrams, ${\cal E}^{(f')}$,
and are assumed finite.  
Loops connecting the $\beta$ eikonal with spectator lines are kept fixed 
in the eikonal field, with minus momenta at order $m^2/Q$.

Now when a subset of active plus loop momenta 
flowing between the $\beta$ eikonal and the active loops has $q^+ \sim Q$,
any soft gluons factorize from the resulting larger $\beta$ jet in the normal fashion
\cite{Collins:1988ig,Bodwin:1984hc,SCfact}.
An example is gluon $k$ of Figs.\ \ref{figs}a, b and c when $q^+\sim Q$.
These contributions are removed by subtractions for the $\beta$ eikonal jet.   
This leaves only the case 
when all active loops have plus momenta of order $m$ or less,
and in this case
 the minus momenta of these loops can be deformed to order  $m$ or more from the origin, away from
poles of the $\beta$ eikonal lines.
But then spectator exchanges with minus momenta
at order $m^2/Q$ are suppressed, unless
they flow into the $\beta$ eikonal before the earliest active-exchange loop. 
For example, with $q^+\le m$ in Fig.\ \ref{figs}, Figs.\ \ref{figs}b and \ref{figs}c are suppressed, 
while only \ref{figs}a remains leading
in this region.
 Indeed, any diagram in which a spectator exchange connects to the
$\beta$ eikonal  between an active exchange and the vertex at $x^-=0$ vanishes 
up to a power of $Q$ after 
integration of the active-exchange minus momentum \cite{AS}.    

This reasoning results in
a decoupling of spectator from active exchange, so that
we may represent the subtracted form of ${\cal M}$ as\bea
&& \left [{\cal M}(p_a,\{p_t\})\right]^{\rm (Sub)} = 
\Big \langle 0 \Big | \Phi^{(f')}(0,-\infty) \bar C_{f'f}(p_a)
\nonumber\\
&& \hspace{0mm} \times\ \prod_t \int d^2{\bf x}_t\,  e^{i{\bf p}_{t,\perp}\cdot {\bf x}_t}\, 
W_-^{(t)}({\bf x}_t)
\Big | 0 \Big \rangle \Psi_f(p_a,\{{p^-}_t,{\bf x}_t\}))\, ,
\nonumber\\
&& \hspace{0mm}\bar C_{f'f}
=  \Big \langle 0 \Big | \Phi^{(f')}(0,-\infty)\, C_{f'f}
\Phi^{(f)}(0,-\infty) \Big |0 \Big \rangle^{\rm (Sub)}\, ,
\nonumber\\
\label{subMjlines}
\eea
where sums over color indices are implicit, and the superscript ``(Sub)"
denotes subtractions.   
The collinear subtractions for each color projection of $\bar C_{f'f}$ 
are equivalent to dividing the timelike form factor for
that color projection by its spacelike version, a result
that follows from the exponentiation properties of the eikonal
form factors \cite{webs}.

We now close the plus integration contours
that flow into spectators  in 
the hybrid form, 
Eq.\ (\ref{Mj}), for Eq.\ (\ref{subMjlines})
in  the lower half-plane, picking up only final-state poles.
The resulting leading regions of ${\cal M}$
are 
characterized either by enlarged $\beta$ jets, which are eliminated
by subtractions, or by multiple
subdiagrams $w_i(q_i)$ that 
 carry vanishing net plus momentum $q_i^+$ into the hard vertex
 at $x^-=0$.   The simplest of the $w_i(q_i)$
are single exchanged gluons, but more generally,
they are generalized {\it webs}, defined by extending
the concept identified for form factors \cite{webs}
as follows.
A subdiagram exchanged between the $\beta$ eikonal and the active
eikonal is a web when it is irreducible under cutting of the two eikonal lines.
Similarly, when a subdiagram is exchanged between the $\beta$ eikonal and 
a spectator eikonal, it is a web if it is irreducible when all connections
to the spectator are in the final state, corresponding to the 
convention that all spectator poles are 
are taken in the lower half plane.

The cancellation of
the eikonal field in Eq.\ (\ref{subMjlines}) 
is now relatively straightforward.
The  contributions of ${\cal E}^{(f')}$, associated
with the spectator-exchange webs $w_i$, cancel by moving the (real) webs one by one
from the amplitude in (\ref{subMjlines}) to the complex conjugate, as in
the eikonal identity above.   
The cancellation of final state interactions ensures that color
factors remain the same in each term.
Once the spectator-exchange webs have been eliminated,  the factors
$\bar C_{f'f}$ and their complex
conjugates are neighboring factors in color traces.
The  subtracted eikonal form factor $\bar C_{f'f}$ of Eq.\ (\ref{subMjlines}) for each
color representation is
a pure phase.  These phases then cancel, 
because, as noted above, the color representation for $\bar C_{f'f}$ is the same
in the amplitude and the complex conjugate after the cancellation of final state interactions.

 As a simple example, consider ${\cal M}(p_a)$ in an abelian theory with coupling $\alpha$.
 Taking for simplicity a single spectator of momentum $p_s$, with
unit charge and with a neutral initial incoming  state, we find from 
  Eq. (\ref{subMjlines}),
\bea
\left[{\cal M}(p_s^-,b)\right]^{\rm (Sub)} &\propto& \ \Psi_f(p_s^-,{\bf b}) 
\nonumber\\
&\ & \hspace{-15mm}
\times
\exp \left( i\frac{\alpha}{2\pi} 
\int \frac{d^2q_\perp}{q_\perp^2}\left[e^{i{\bf b}\cdot q_\perp} -1\right] \right)\, ,
\label{Mj2}
\eea
in which the phase is explicit.

{\bf Weights, Transverse Scales and Phases.} 
Our discussion above assumes a single scale $m$ for transverse momenta,
which sets the off-shellness of virtual lines at order $m^2$.   
For subdiagrams whose lines have, for example, much lower transverse momenta,
$m'\ll m$, lines at order $m$ may be considered as part of the hard scattering.
	  By the power counting of Refs.\ \cite{Libby:1978nr,Sterman:1978bi},
we can order sets of spectators according to their transverse momenta,
and, in effect, can start with lines of the lowest transverse momenta and
``work our way in" to the hard scattering.   Softer, non-collinear lines can
attach ladders of the $p$ jet with differing transverse momenta only to final state partons
and to the $\beta$ eikonal,
and not other ladders.  Such final state soft gluon connections cancel in sums over
final states.   Eikonal field exchange gluons are treated as above for each scale $m$.
 
Our results on soft-gluon cancellation depend 
crucially on an inclusive sum over final states.   Only after the cancellation
of final states can we move spectator-connected webs across the final
state, canceling the phases that they generate in Eq.\ (\ref{Mjlines}).   
This cancellation will persist only to scales where soft radiation emitted from spectator
lines into the final state is fully summed over.   For fixed 
transverse scale $m$, 
wide-angle
final state radiation from spectators is characterized
by energies of order $m^2/Q$, with $Q$ the 
spectator momentum.   So long as 
the weight $S(N)$ is chosen so that radiation of this scale is summed
inclusively {\it in all directions}, 
soft logarithms will cancel
in the transition probability.   Such weighted ${\cal S}$ are thus infrared
finite, and correspondingly, cross sections defined with 
these weights
factorize.
At the same time, if a weighted cross section does not allow radiation 
greater than energy $E_0$ in 
{\it any} angular region at wide angles,
we may anticipate 
the noncancellation of radiation at that energy scale, leading to 
logarithms of the ratio $E_0/Q$ at each order, whether from spectators, or from
high-$p_T$ partons in directions 
allowed
by $S(N)$ \cite{Dasgupta:2001sh}.   At sufficiently high orders,
the effects of
phases due to emission from spectators 
of transverse momentum $\sqrt{E_0Q}$ or greater also 
begin to appear in finite logarithmic corrections to the cross section, with enhancements
associated with the emission of those spectators at ordered
transverse momentum
 \cite{Forshaw:2006fk}.   
 
 In summary, we have outlined arguments for
  the cancellation of infrared singularities 
in transition probabilities with colored incoming lines and final-state
jets.   We have also seen that the pattern
of cancellation  is related to the generation of phases in hard-scattering amplitudes,
the exchange of Glauber gluons,
and the flow of energy into final states.   
Further investigation 
should shed more
light on these important issues, and their implications for phenomenology.

\acknowledgments
The work of 
MA and GS was supported in part by the National Science Foundation, 
grants PHY-0354776, PHY-0354822 and PHY-0653342,
and of MA by Swiss National Science Foundation contract number 200021-117873.
We thank Michael Seymour and Ilmo Sung for useful conversations.

\end{document}